\documentclass{webofc}

\usepackage[varg]{txfonts}   
\usepackage{hyperref}
\usepackage{url}
\hypersetup{colorlinks=true,citecolor=blue,urlcolor=blue,linkcolor=blue}
\begin{document}
\title{Extraction of jet-medium interaction details through jet substructure for inclusive and gamma-tagged jets}
%
%

\author{\firstname{Yasuki} \lastname{Tachibana}\inst{1}\fnsep\thanks{\email{ytachibana@aiu.ac.jp}} 
for the JETSCAPE Collaboration
}



\institute{Akita International University, Yuwa, Akita-city 010-1292, Japan}

\abstract{%
We present a comprehensive study of jet substructure modifications in high-energy heavy-ion collisions using both inclusive jets and $\gamma$-tagged jets, based on a multi-stage jet evolution model within the Monte Carlo framework JETSCAPE. To investigate hard parton splittings inside jets, we focus on Soft Drop observables. Our results for the groomed splitting radius and groomed jet mass distributions of inclusive jets show a slight narrowing compared to proton-proton baselines. We demonstrate that this apparent narrowing is primarily a selection bias from energy loss, rather than a direct modification of the splitting structure, by analyzing $\gamma$-tagged jets, where such bias is eliminated or significantly reduced.
We also show that quark jets exhibit genuine modifications in their splitting structure, which is not seen in gluon jets. These effects are clearly visible in the substructure of $\gamma$-tagged jets, which are dominated by quark jets, but are not apparent for inclusive jets. This demonstrates that $\gamma$-tagged jets offer a powerful probe of medium-induced modifications to the hard splitting structure of jets. 
}

\maketitle
\section{Introduction}
\label{intro}
The Soft Drop grooming algorithm~\cite{Larkoski:2014wba} provides access to the hard parton splittings within a reconstructed jet. The observables obtained through grooming are relatively well described by perturbative calculations, making them sensitive probes of medium effects on parton shower evolution in high-energy heavy-ion collisions. However, inclusive jet measurements in such collisions are subject to selection bias: jets that are more susceptible to substantial energy loss in the medium tend to have significantly reduced $p_{\mathrm{T}}^{\mathrm{jet}}$, making them less likely to satisfy the $p_{\mathrm{T}}^{\mathrm{jet}}$ trigger condition. This can obscure the true impact of jet quenching on jet substructure. To mitigate this bias, we consider $\gamma$-tagged jets, where the recoiling photon provides an approximate reference for the $p_{\mathrm{T}}^{\mathrm{jet}}$ value in the absence of medium-induced modifications. Furthermore, since $\gamma$-associated processes predominantly produce quarks, comparing $\gamma$-tagged jets with inclusive jets enables systematic studies of flavor-dependent effects.

In these proceedings, we perform a systematic and comprehensive analysis of jet substructure modifications using the multi-stage jet evolution model implemented in the \textsc{jetscape} framework. By investigating the effects of selection bias and flavor dependence, we demonstrate that $\gamma$-tagged jets serve as a clear probe for accessing detailed information on medium-induced modifications to parton splittings in jets.

\section{Model and Simulations}
\label{model}
We perform simulations using the multi-scale in-medium jet evolution model \textsc{matter}+\textsc{lbt} of JETSCAPEv3.5 AA22 tune, implemented within the \textsc{jetscape} framework
~\cite{JETSCAPE:2022jer, JETSCAPE:2022hcb, JETSCAPE:2023hqn, JETSCAPE:2025rip}. The \textsc{matter} module~\cite{Majumder:2013re,Cao:2017qpx} describes high-virtuality partons undergoing vacuum-like showers with strong suppression from modified coherence effects~\cite{Kumar:2019uvu,Kumar:2025rsa}, while the \textsc{lbt} module~\cite{Wang:2013cia,He:2015pra,Cao:2016gvr}  handles low-virtuality partons based on kinetic theory with the on-shell approximation.  
In this study, we simulate jet events in PbPb collisions at $\sqrt{s_{\mathrm{NN}}} = 5.02~\mathrm{TeV}$. The space-time medium profile is obtained from a $(2+1)$-dimensional freestreaming evolution~\cite{Liu:2015nwa} followed by viscous hydrodynamics using \textsc{vishnu}~\cite{Shen:2014vra}, with the \textsc{trento} initial conditions~\cite{Moreland:2014oya}. 

\section{Results}
\label{results}
We present results from the \textsc{matter}+\textsc{lbt} within the \textsc{jetscape} framework for the groomed splitting radius $r_{\mathrm{g}}$ and groomed jet mass $m_{\mathrm{g}}$ distributions, which are sensitive to the hard parton splittings inside jets. 
Throughout this study, the medium profile from 0--10\% central Pb$+$Pb collisions is employed, and the Soft Drop parameters are fixed to $z_{\mathrm{cut}}=0.2$ and $\beta = 0$.

\subsection{Single-parton-initiated jet shower simulation}
\label{pgun}
To facilitate systematic studies under controlled conditions, we present results from the \textsc{matter}$+$\textsc{lbt} simulations of events with a single high-energy parton propagating at midrapidity in a random azimuthal direction. 
The species of the parent parton, either a gluon or a massless light quark, and its initial energy, $E_{\mathrm{init}} = 140$ GeV, are fixed. 

\begin{figure}[h]
\centering
\includegraphics[width=0.9\textwidth,clip]{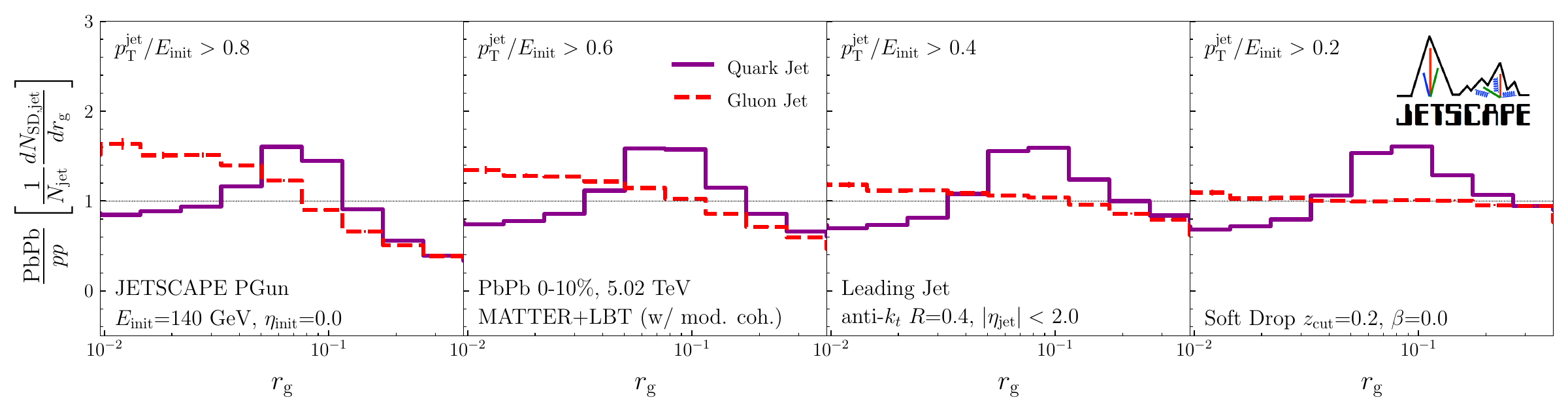}
\includegraphics[width=0.9\textwidth,clip]{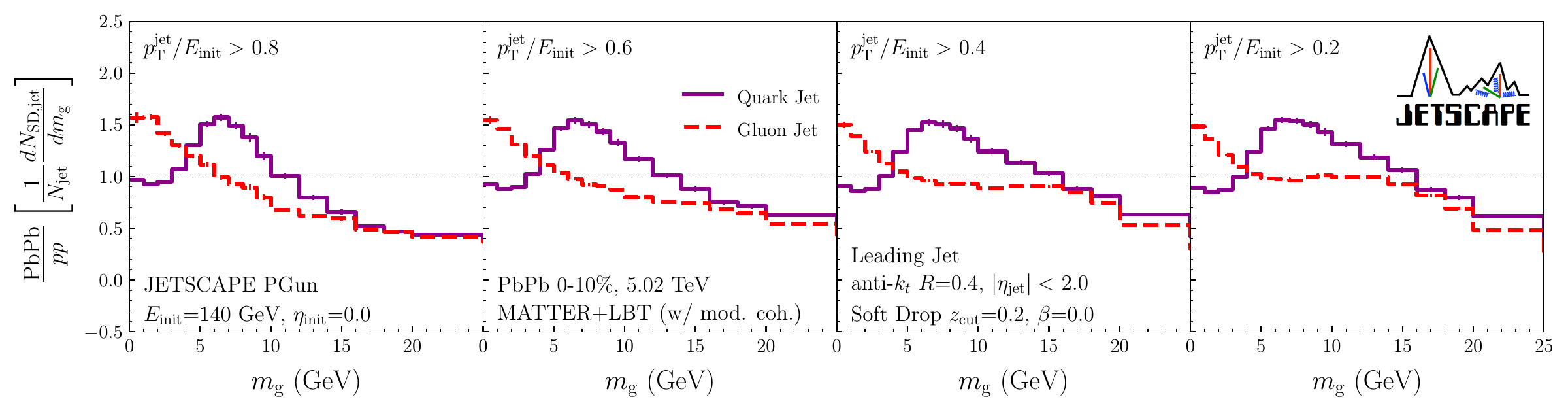}
\caption{%
Ratios of the $r_{\mathrm{g}}$ (top) and $m_{\mathrm{g}}$ (bottom) distributions for single-parton-initiated jets propagating through the QGP medium produced in 0--10\% central Pb$+$Pb collisions, relative to those in vacuum. Results are shown for quark-initiated (solid) and gluon-initiated (dashed) jets, with the initial parton energy fixed at $E_{\mathrm{init}} = 140$ GeV, under different $p_{\mathrm{T}}^{\mathrm{jet}}$ trigger thresholds: 112, 84, 56, and 28 GeV. %
}
\label{fig-1}       
\end{figure}

Figure~\ref{fig-1} shows the modification of the distributions of $r_{\mathrm{g}}$ and $m_{\mathrm{g}}$ from Soft Drop. 
A bump indicative of jet hard splitting broadening is observed in quark jets in both the $r_{\mathrm{g}}$ and $m_{\mathrm{g}}$ distributions, whereas no such feature appears in gluon jets. This behavior arises from the initially narrow structure of quark jets, which allows medium effects to induce wider-angle hard splittings than those occurring during the early vacuum-like stage. In contrast, the hard, vacuum-like branchings of gluon jets remain largely unaffected by the medium. 

Reducing the $p_{\mathrm{T}}^{\mathrm{jet}}$ trigger thresholds mitigates the effect of selection bias, and the suppression at large $r_{\mathrm{g}}$ disappears. This clearly indicates that the suppression observed at large $r_{\mathrm{g}}$ with higher $p_{\mathrm{T}}^{\mathrm{jet}}$ thresholds originates from the selection bias. 
For the $m_{\mathrm{g}}$ distribution, the selection bias cannot be removed simply by varying the $p_{\mathrm{T}}^{\mathrm{jet}}$ trigger threshold. This is because medium-induced large-angle emissions are often removed in Soft Drop or the jet reconstruction, leading to a loss in jet mass. As a result, the suppression at large $m_{\mathrm{g}}$ persists.

\subsection{Prediction for inclusive jets and $\gamma$-tagged jets}
\label{pythia}
Finally, we present the \textsc{matter}$+$\textsc{lbt} predictions from simulations incorporating realistic hard scattering events generated with \textsc{pythia}~8, including initial-state radiation and multi-parton interactions, for both inclusive and $\gamma$-tagged jets.
For $\gamma$-tagged jets, statistical precision is enhanced by selectively generating events in which the photon originates from leading-order hard scattering processes in \textsc{pythia}~8. 

\begin{figure}[h]
\centering
\includegraphics[width=0.9\textwidth,clip]{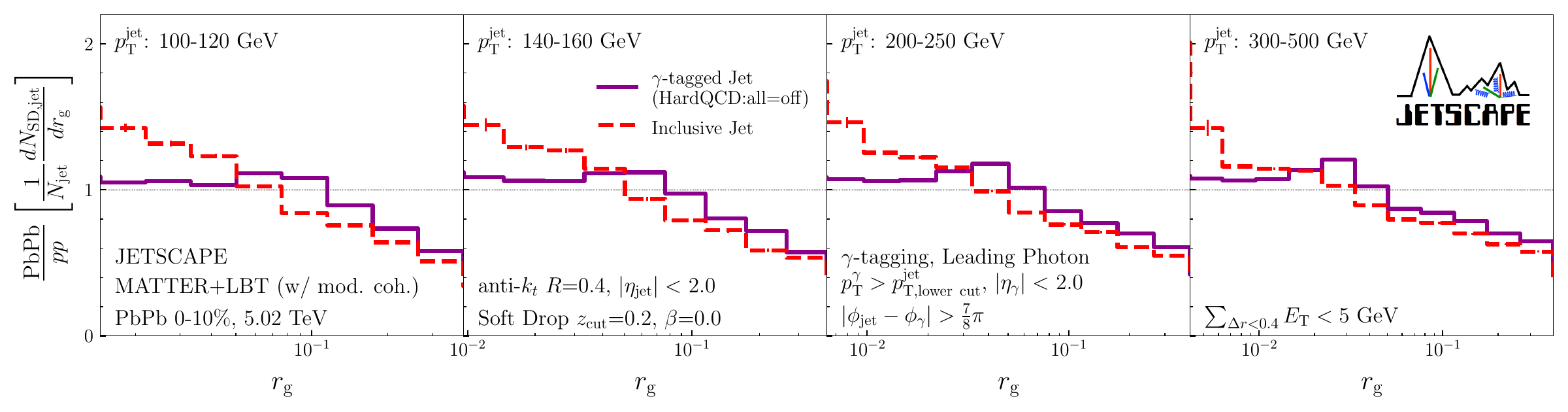}
\includegraphics[width=0.9\textwidth,clip]{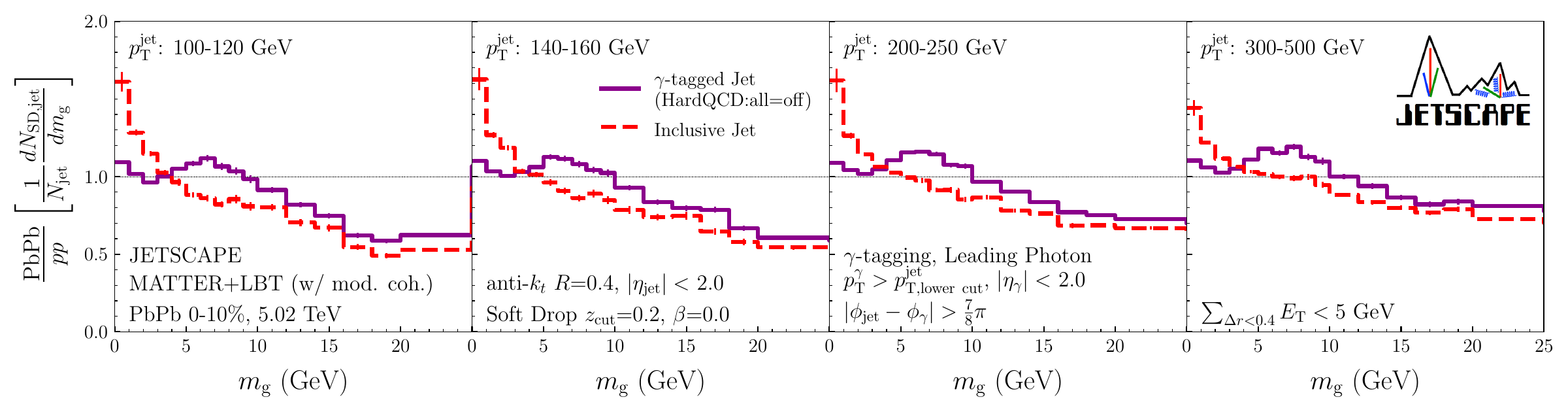}
\caption{%
Ratios of the $r_{\mathrm{g}}$ (top) and $m_{\mathrm{g}}$ (bottom) distributions for inclusive jets (solid) and $\gamma$-tagged jets (dashed) in 0--10\% central Pb$+$Pb collisions, relative to those in p$+$p collisions, are shown for different $p_{\mathrm{T}}^{\mathrm{jet}}$ ranges.
}
\label{fig-2}       
\end{figure}

Figure~\ref{fig-2} shows the medium-induced modifications of the $r_{\mathrm{g}}$ and $m_{\mathrm{g}}$ distributions for inclusive and $\gamma$-tagged jets, triggered by different $p_{\mathrm{T}}^{\mathrm{jet}}$ thresholds.
Inclusive jets, which are predominantly gluon-initiated, exhibit modifications consistent with gluon jet behavior: a monotonic narrowing of the groomed splitting angle ($r_{\mathrm{g}}$) and a systematic shift toward lower groomed masses ($m_{\mathrm{g}}$), both driven by selection bias.
In contrast, although $\gamma$-tagged jets are also subject to some degree of selection-bias-induced narrowing, they exhibit a pronounced bump structure in both the $r_{\mathrm{g}}$ and $m_{\mathrm{g}}$ distributions. This feature reflects significant medium-induced broadening of the hard splitting, characteristic of quark-initiated jets.

\begin{figure}[h]
\centering
\includegraphics[width=0.9\textwidth,clip]{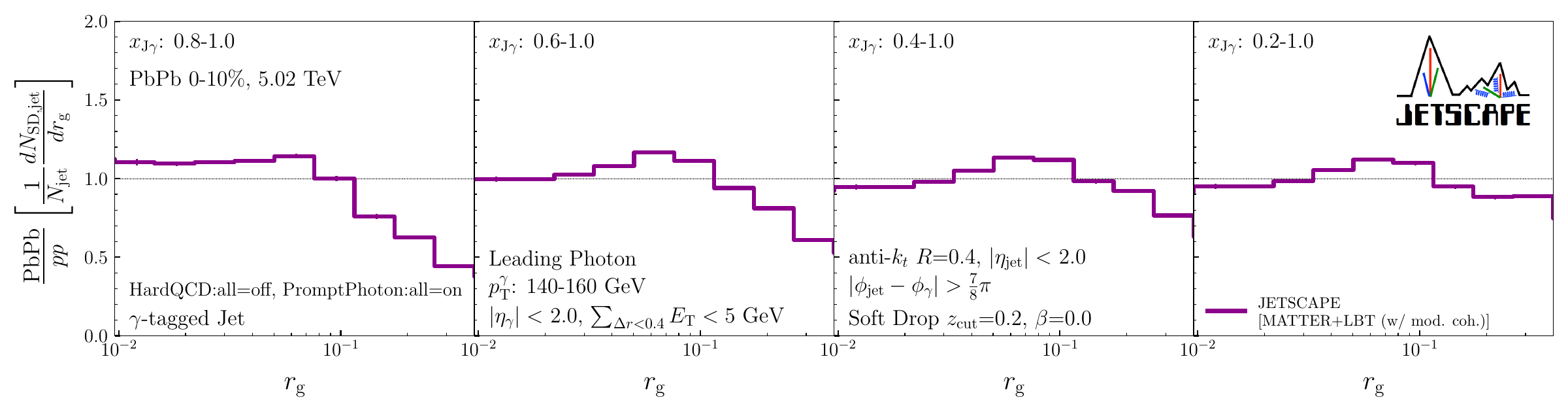}
\includegraphics[width=0.9\textwidth,clip]{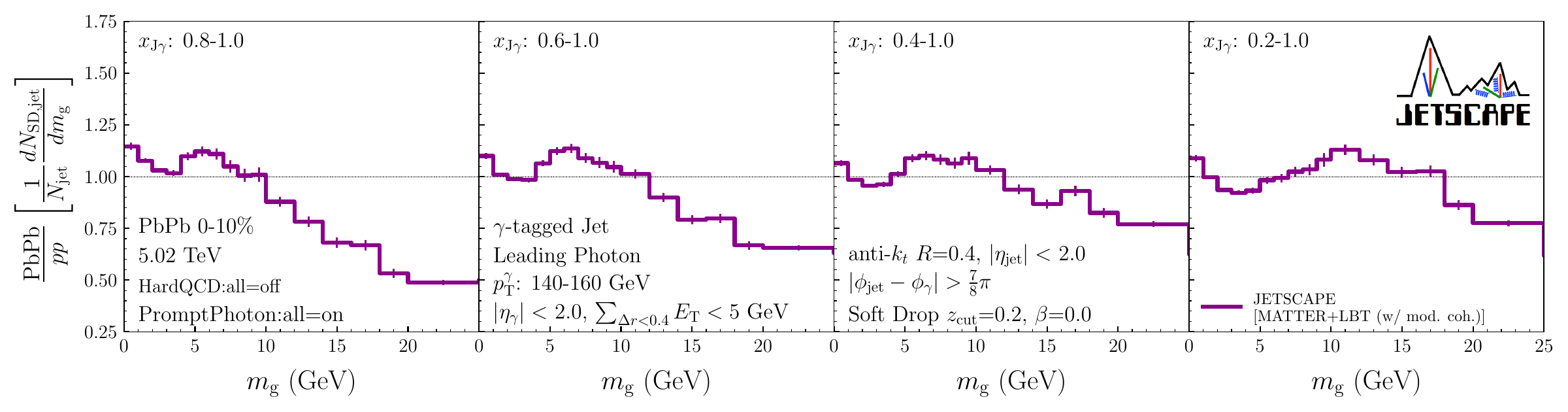}
\caption{%
Ratios of the $r_{\mathrm{g}}$ (top) and $m_{\mathrm{g}}$ (bottom) distributions for $\gamma$-tagged jets in 0--10\% central Pb$+$Pb collisions, relative to those in p$+$p collisions, are shown for different $x_{\mathrm{J\gamma}}=p_{\mathrm{T}}^{\mathrm{jet}}/p_{\mathrm{T}}^{\gamma}$ trigger thresholds with fixed $p_{\mathrm{T}}^{\gamma}$ threshold ($140<p_{\mathrm{T}}^{\gamma}<160$~GeV). 
}
\label{fig-3}       
\end{figure}

Figure~\ref{fig-3} illustrates the impact of the $p_{\mathrm{T}}^{\mathrm{jet}}$ trigger threshold, expressed as $x_{\mathrm{J\gamma}} = p_{\mathrm{T}}^{\mathrm{jet}} / p_{\mathrm{T}}^{\gamma}$, on the $r_{\mathrm{g}}$ and $m_{\mathrm{g}}$ distributions for $\gamma$-tagged jets, with $p_{\mathrm{T}}^{\gamma}$ held fixed. 
A distinctive feature of quark-dominated jets is the emergence of a broadening-induced bump in both observables.
As the lower bound of the $x_{\mathrm{J\gamma}}$ cut is reduced, jets with substantial energy loss are retained in the sample. This mitigates the selection bias in the $r_{\mathrm{g}}$ distribution, recovering the suppressed large-$r_{\mathrm{g}}$ region.
However, as demonstrated in single-jet simulations, even when the $x_{\mathrm{J\gamma}}$ cut is relaxed, due to the wide-angle soft radiation, the suppression at large $m_{\mathrm{g}}$ persists. 

\section{Conclusion}
\label{conclusion}
We have performed a systematic analysis of the modification of Soft Drop jet substructure in central Pb$+$Pb collisions at $\sqrt{s_{\mathrm{NN}}}=5.02$~TeV using the \textsc{MATTER}$+$\textsc{LBT} model within the \textsc{JETSCAPE} framework.
Inclusive jets exhibit a monotonic narrowing of the groomed splitting radius $r_{\mathrm{g}}$ and an overall reduction in groomed mass $m_{\mathrm{g}}$ relative to proton-proton baselines.
Controlled simulations with single-parton-initiated jets reveal that this behavior is primarily driven by selection bias: jets that undergo greater energy loss---and therefore tend to have wider splittings or larger mass---often fall below the $p^{\mathrm{jet}}_{\mathrm{T}}$ threshold.

In contrast, the $r_{\mathrm{g}}$ and $m_{\mathrm{g}}$ distributions for $\gamma$-tagged jets exhibit pronounced bumps, indicative of significant medium-induced broadening of the hard splitting, which is a characteristic feature of quark-initiated jets.
Moreover, $\gamma$-tagged jets enable a systematic investigation of selection bias effects present in inclusive jet measurements. For instance, by varying the $x_{\mathrm{J}\gamma}$ cut, one can analyze jets with different levels of energy loss.
In the case of $r_{\mathrm{g}}$, the narrowing observed in inclusive jets vanishes when jets with low $x_{\mathrm{J}\gamma}$ are included, confirming that the apparent narrowing is entirely attributable to selection bias.
In contrast, for $m_{\mathrm{g}}$, such a correction is not feasible due to the partial loss of jet energy during reconstruction and grooming. The Soft Drop procedure, in particular, removes soft radiation, resulting in a reduction in mass.  
These findings establish $\gamma$-tagged, groomed observables as powerful probes to precisely elucidate the nature of jet-medium interactions.

\section*{Acknowledgments}
\label{ack}
This work was supported in part by the National Science Foundation (NSF) within the framework of the JETSCAPE collaboration, under grant number OAC-2004571 (CSSI:X-SCAPE), 
the U.S. Department of Energy (DOE) under grant number DE-SC0013460, the Office of the Vice President for Research (OVPR) at Wayne State University, and JSPS KAKENHI Grant Numbers~22K14041 and 25K07303.

%
\bibliography{template.bib} 
%
%

\end{document}